\documentclass{iau} 
\usepackage{graphicx}

\title[The hosts of early ionised bubbles] 
{Unveiling the most luminous Lyman-$\alpha$ emitters in the epoch of reionisation}

\author[Matthee \& Sobral]   
{Jorryt Matthee$^1$ \and David Sobral$^2$}

\affiliation{$^1$Department of Physics, ETH Z\"urich, Wolfgang-Pauli-Strasse 27, 8093 Z\"urich, Switzerland \\ email: {\tt mattheej@phys.ethz.ch} \\[\affilskip]
$^2$ Department of Physics, Lancaster University, Lancaster, LA1 4YB, United Kingdom \\
}

\pubyear{2019}
\volume{352}  
\jname{Uncovering early galaxy evolution in the ALMA and JWST era}
\editors{}
\begin{document}

\maketitle

\begin{abstract}
Distant luminous Lyman-$\alpha$ emitters are excellent targets for detailed observations of galaxies in the epoch of reionisation. Spatially resolved observations of these galaxies allow us to simultaneously probe the emission from young stars, partially ionised gas in the interstellar medium and to constrain the properties of the surrounding hydrogen in the circumgalactic medium. We review recent results from (spectroscopic) follow-up studies of the rest-frame UV, Lyman-$\alpha$ and [CII] emission in luminous galaxies observed $\sim500$ Myr after the Big Bang with ALMA, HST/WFC3 and VLT/X-SHOOTER. These galaxies likely reside in early ionised bubbles and are complex systems, consisting of multiple well separated and resolved components where traces of metals are already present.
\end{abstract}

\firstsection 
\section{Introduction}
Thanks to its rest-frame UV wavelength and intrinsic brightness in star-forming regions, the Lyman-$\alpha$ (Ly$\alpha$, $\lambda_0=1215.67$ {\AA}) emission line is extremely useful for identifying galaxies in the early Universe ($z>2$). In recent years, significant progress has been made in enlarging the dynamical range in the luminosities of known Lyman$-\alpha$ emitters (LAEs): either by identifying samples of very luminous LAEs using wide-field narrow-band surveys (e.g. \cite[Matthee et al. 2015; Zheng et al. 2017; Konno et al. 2018; Sobral et al. 2018]{Matthee2015,Zheng2017,Konno2018,Sobral2018}) or by identifying the faintest LAEs using sensitive integral field spectrographs such as MUSE (e.g. \cite[Bacon et al. 2017]{Bacon2017}).

Due to the high scattering cross-section at line-centre, Ly$\alpha$ photons are subject to radiative transfer effects in the presence of neutral hydrogen, which results in an uncertain Ly$\alpha$ escape fraction. This significantly challenges the interpretation of observed Ly$\alpha$ light. However, the sensitivity of Ly$\alpha$ to the HI column density also provides an opportunity to study the HI structure in and around galaxies. This is of particular relevance for the escape of ionising photons from the ISM (\cite[e.g. Verhamme et al. 2015]{Verhamme2015}) and for using Ly$\alpha$ emission as a tracer of reionisation (\cite[e.g. Mason et al. 2018]{Mason2018}).

Here we present highlights on results obtained using spectroscopic observations of luminous LAEs at $z\approx6.5$ (selected in \cite[Matthee et al. 2015; Sobral et al. 2015]{Matthee2015,Sobral2015}). While all galaxies in the sample have a high Ly$\alpha$ luminosity, they span a range in UV continuum luminosity (1-5 M$_{1500}^{\star}$) and a range in Ly$\alpha$ equivalent width (EW$_0$=30-200 {\AA}). Current facilities such as ALMA and VLT/X-SHOOTER can be used to obtain a first glimpse of the properties of the ISM in these galaxies in the first billion year of cosmic time. 

In \S 2 we focus on recent results on the LAE `COLA1' (\cite[Hu et al. 2016; Matthee et al. 2018]{Hu2016,Matthee2018}), a unique double peaked Ly$\alpha$ emitter at $z=6.59$ that is of particular interest for the study of the epoch of reionization. In \S 3 we review results of recent ALMA follow-up observations of Ly$\alpha$-selected galaxies at $z\approx6-7$. We use a $\Lambda$CDM cosmology with $\Omega_{\Lambda} = 0.70$, $\Omega_{\rm M} = 0.30$ and H$_0 = 70$ km s$^{-1}$ Mpc$^{-1}$ and a \cite{Salpeter1955} IMF.

\begin{figure}[t]
\begin{tabular}{cc}
 \includegraphics[height=5.65cm]{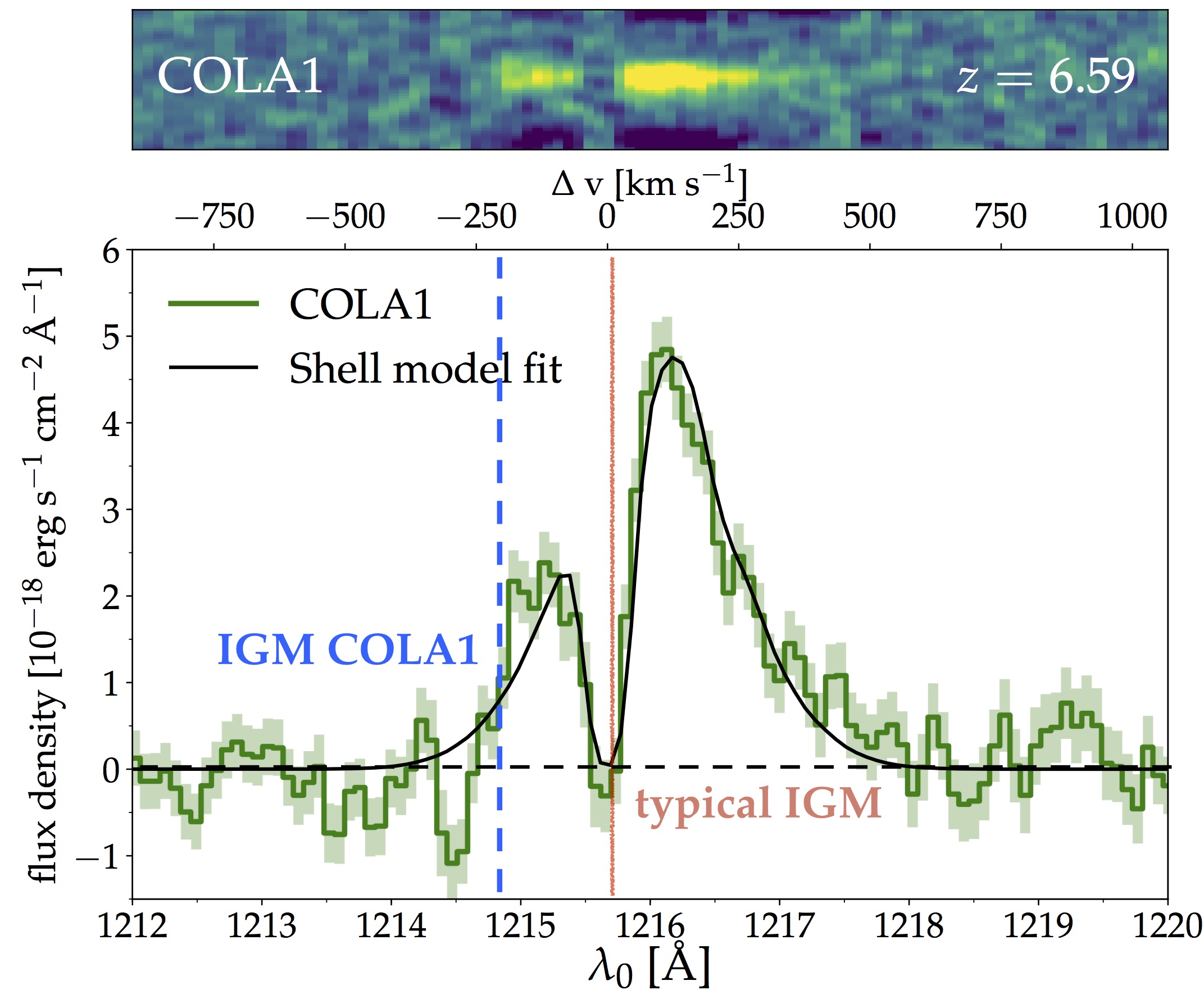} &
 \includegraphics[height=5.57cm]{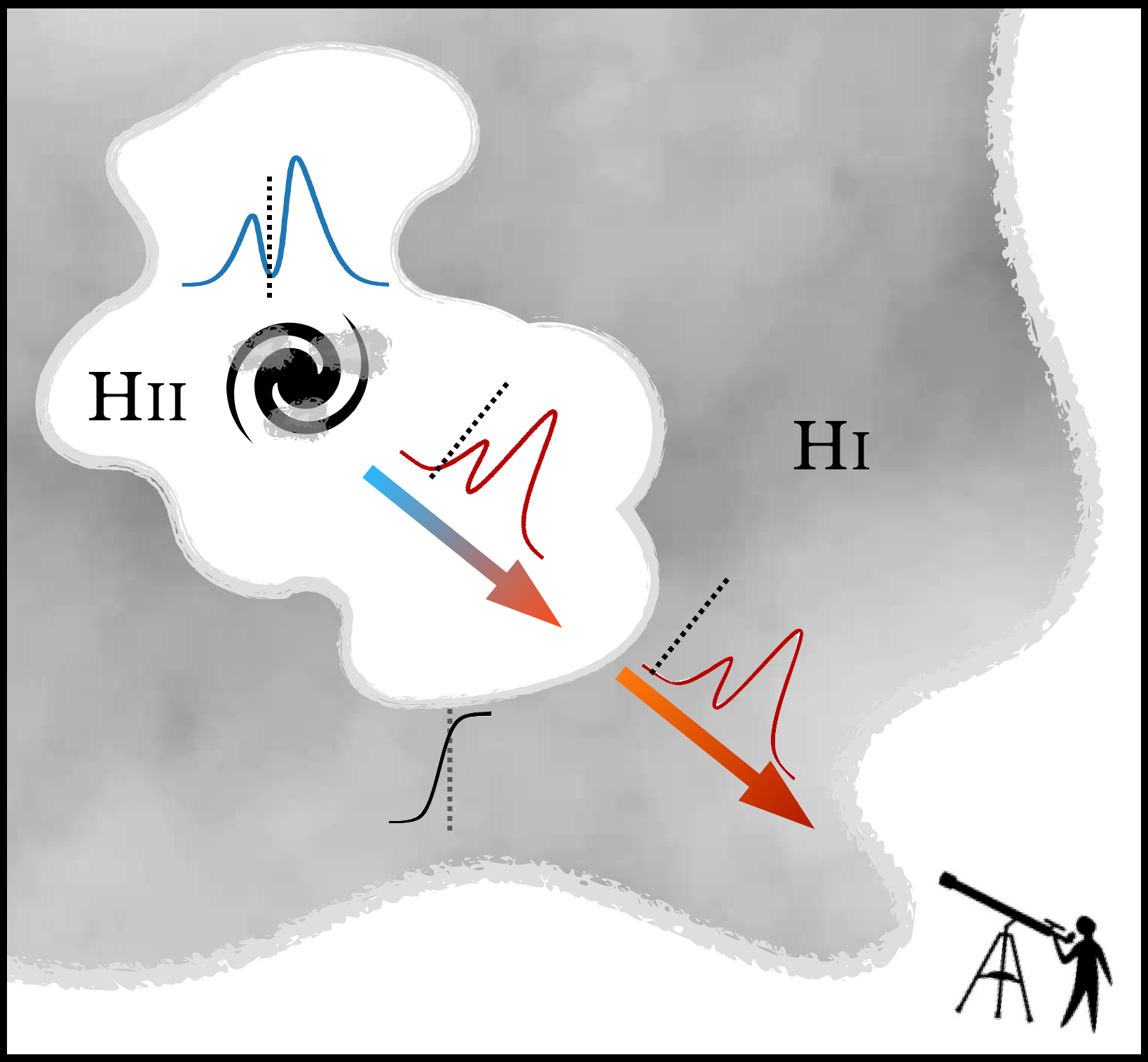}
 \end{tabular}
 \caption{Left: VLT/X-SHOOTER spectrum of the double peaked Ly$\alpha$ line in COLA1 at $z=6.591$ (adapted from \cite[Matthee et al. 2018]{Matthee2018}).The red line indicates the typical velocity at which the Ly$\alpha$ spectrum in typical LAEs is truncated, likely due to low IGM transmission. The blue dashed line indicates the velocity at which this seems to occur for COLA1. Right: Illustration of a scenario in which COLA1 resides in a large ionised bubble. Ly$\alpha$ redshift out of the resonant frequency on the Hubble flow before encountering significant amounts of neutral hydrogen. This allows the blue peak to be observed.}
   \label{fig1}
\end{figure}

\section{COLA1: a double peaked Ly$\alpha$ emitter at $z=6.59$} \label{COLA1}
At high redshift ($z>6$) virtually all LAEs show a single red asymmetric peak (e.g. \cite[Hu et al. 2010; Matthee et al. 2017b]{Hu2010,Matthee2017SPEC}), likely because the increasing hydrogen density around galaxies results in a low transmission on the blue side of line-center (\cite[e.g. Laursen et al. 2011]{Laursen2011}). The discovery of double peaked Ly$\alpha$ emission at $z=6.59$ in `COLA1' by \cite{Hu2016} challenged this picture. 

Initially, the reality of COLA1 being a LAE at $z=6.59$ was uncertain. The peak separation is consistent with the line being the [OII] doublet at $z=1.47$ and tentative optical flux has been observed in bands blue-wards of the Lyman-break at $z=6.6$ (\cite[Matthee et al. 2018]{Matthee2018}). However, new higher resolution, deeper VLT/X-SHOOTER observations (left panel of Fig.\,\ref{fig1}; \cite[Matthee et al. 2018]{Matthee2018}) rule out that COLA1 is an [OII] emitter at $z=1.47$ based on 1) the asymmetry of the red line, 2) the zero-flux between the lines,  3) the ratio of the blue-to-red line, 4) the non-detection of H$\alpha$ at $z=1.47$ and 5) the extreme observed EW. Moreover, the spectrum also revealed foreground Ly$\alpha$ and [OIII]$_{5008}$ emission at $z=2.14$ at a close ($<1''$) separation that explains the flux in blue optical bands. Therefore, COLA1 is confirmed as a double peaked LAE at $z=6.59$.

How is it possible that double peaked Ly$\alpha$ emission is seen in COLA1? The answer likely lies in the narrowness of the Ly$\alpha$ peak separation ($220$ km s$^{-1}$). Using UV observations of green pea galaxies at $z\approx0.3$ with {\it HST}, \cite{Izotov2018} recently found a strong anti-correlation between the Lyman-continuum escape fraction (f$_{\rm esc, LyC}$) and the peak separation of the double peaked Ly$\alpha$ line. The physical explanation is that LAEs with narrow peak separation have an ISM with low HI column density channels through which ionising photons escape (\cite[Verhamme et al. 2015]{Verhamme2015}). Following the trend from \cite[Izotov et al. 2018]{Izotov2018}), the peak separation in COLA1 corresponds to a high f$_{\rm esc,LyC} \approx30$ \%.

In order to observe double peaked Ly$\alpha$ emission at $z>6$, galaxies need to reside in ionised regions that are large enough to allow Ly$\alpha$ photons to redshift out of the resonance wavelength (right panel of Fig.\,\ref{fig1}). Smaller regions are thus required for galaxies with smaller Ly$\alpha$ peak separations. Since the escape fraction is anti-correlated with peak separation, galaxies with smaller peak separation will more easily reside in large ionised regions. Such galaxies are therefore likely more easily observed at high redshift.

Assuming that the systemic redshift of COLA1 lies between the two Ly$\alpha$ lines (as in lower-redshift double peaked LAEs) and that the blue line needs to redshift on the Hubble flow by 220 km s$^{-1}$ before encountering neutral hydrogen, a Str\"omgren sphere with radius 300 proper kpc is required (\cite[Matthee et al. 2018]{Matthee2018}). Assuming an ionising efficiency $\xi_{ion} = 10^{25.4}$ Hz erg$^{-1}$, an escape fraction of 15 \% (conservative, compared to inferred 30 \%) and a SFR of 30 M$_{\odot}$ yr$^{-1}$ (based on the UV luminosity), COLA1 can plausibly ionise the required region on its own in 10$^7$ yr. 

COLA1 is the first galaxy that can be used as both a tracer and an agent of reionisation. Future observations can assess the properties of the ISM and stellar populations present in COLA1 and investigate whether COLA1 resides in a large over-density. Additionally, detailed statistical investigation of the presence of double peaked Ly$\alpha$ lines (and their peak separations) in galaxies at different UV luminosities and redshifts may provide useful indirect information on which (star-forming) galaxies contribute to the global ionising background throughout the evolution of the Universe.

\section{ALMA results on luminous LAEs at $z\approx6-7$}
Since the advent of the Atacama Large (Sub)Millimetre Array (ALMA), observations have been performed to study the ISM and dust properties of high-redshift galaxies using far-infrared emission lines such as [CII]158$\mu$m and the FIR continuum (IR hereafter). 

\begin{figure}[h]
\begin{tabular}{cc}
\hspace{-0.25cm} \includegraphics[height=5.65cm]{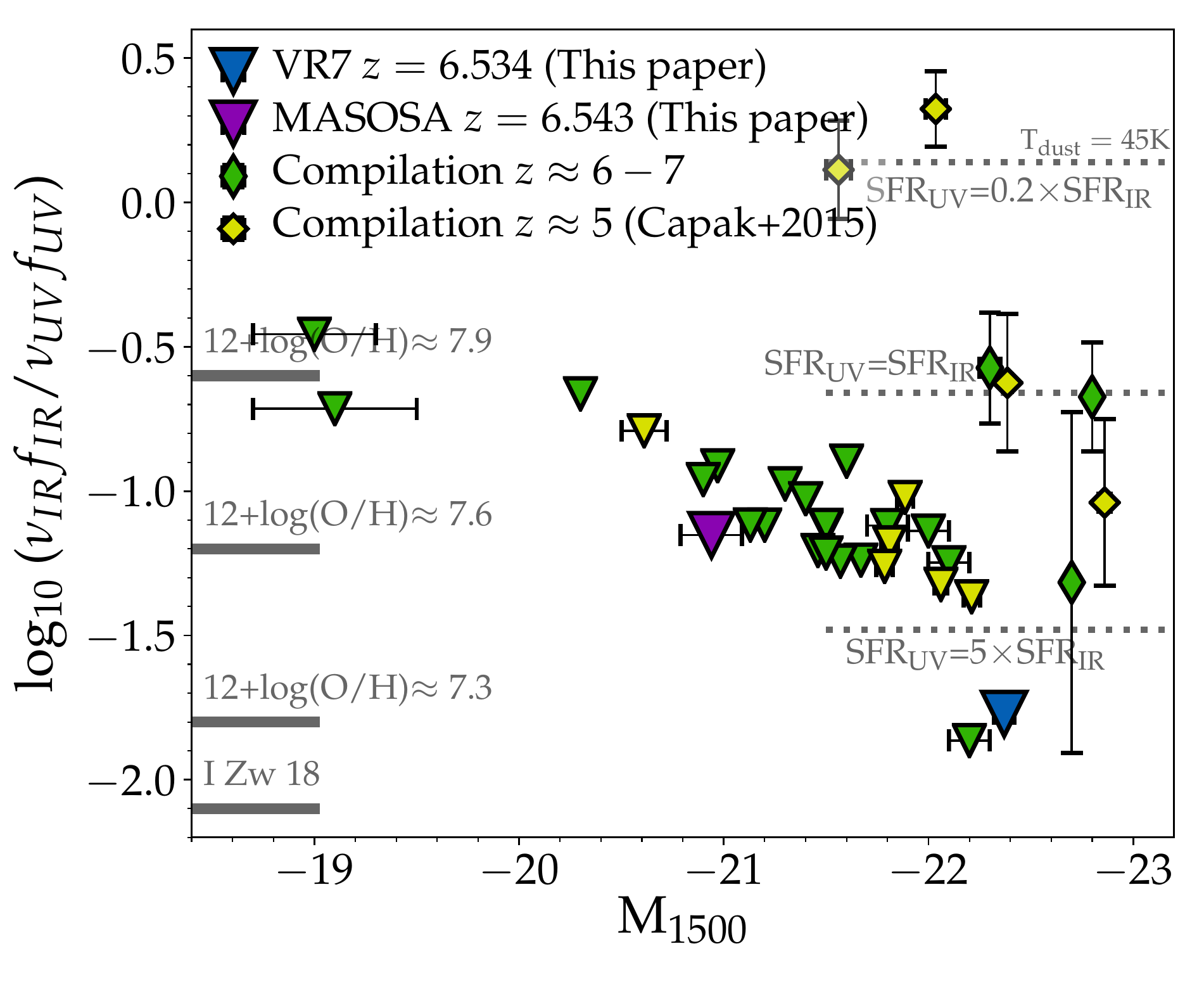}&
\hspace{-0.45cm}  \includegraphics[height=5.65cm]{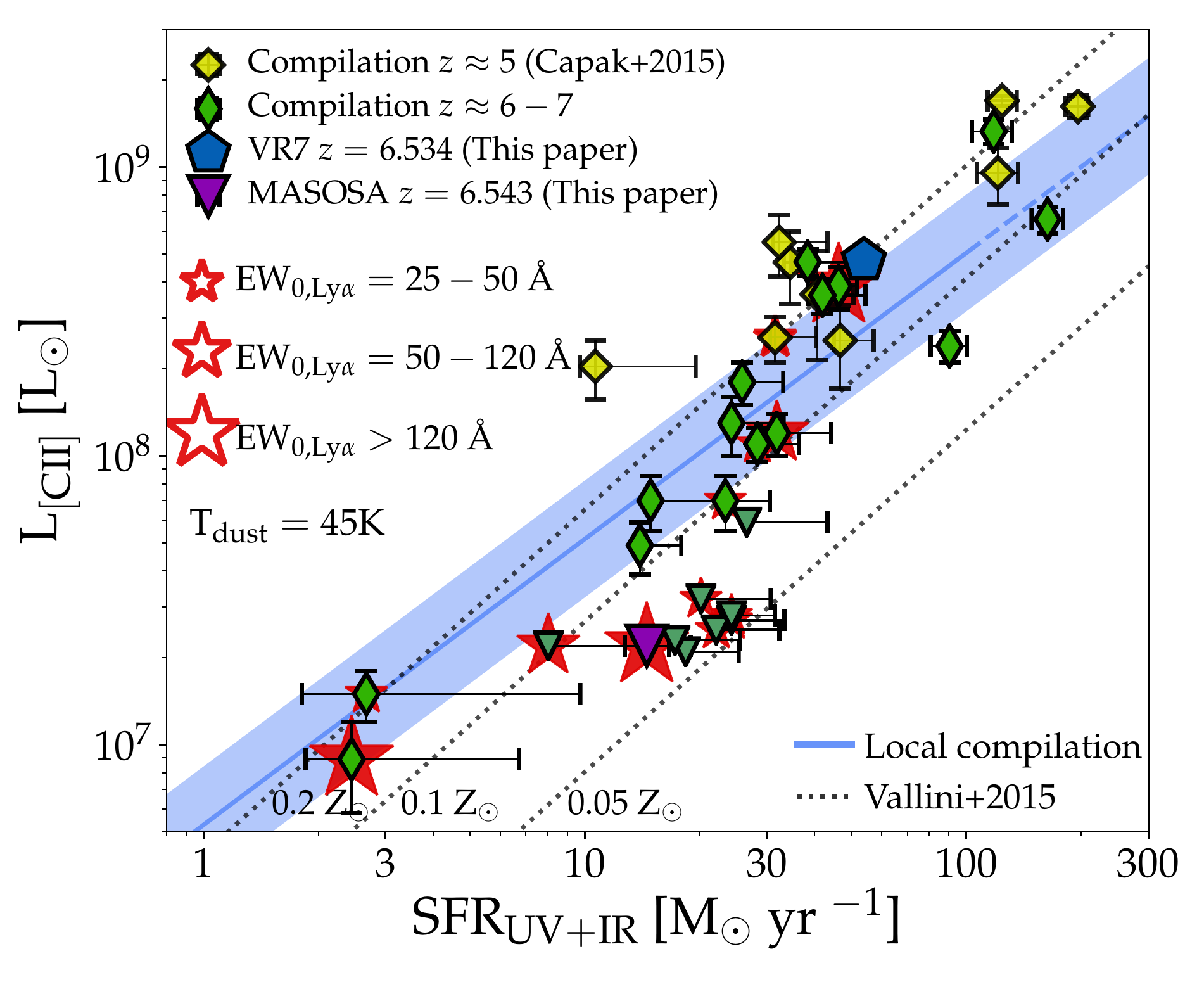}
\end{tabular}
 \caption{Left: Observed rest-frame UV ($\lambda_0 = 1500${\AA}) to IR ($\lambda_0 = 160 \mu$m) flux ratio as a function of UV luminosity for a compilation of UV and Ly$\alpha$-selected galaxies at $z\approx5-7$. The gas-phase metallicity corresponding to galaxies in the local Universe with similar UV to IR flux ratios (\cite[Maiolino et al. 2015]{Maiolino2015}) are illustrated. We also illustrate the obscured to unobscured SFR ratio when converting IR flux to IR luminosity assuming a dust temperature of 45 K. Right: [CII] luminosity versus SFR$_{\rm UV+IR}$ for a compilation of galaxies at $z\approx5-7$ compared to galaxies in the local Universe (blue band). Above SFR$>30$ M$_{\odot}$ yr$^{-1}$, most high-redshift galaxies are relatively luminous in [CII], while low [CII] luminosities (and relatively deep upper limits on the [CII] luminosity) are found in fainter galaxies. Both panels adapted from \cite{Matthee2019}.   }
   \label{fig2}
\end{figure}

As compiled in \cite{Matthee2019}, the number of UV and Ly$\alpha$ selected galaxies with dust continuum detections at $z\approx5-7$ is low. Furthermore, detections/upper limits in a single IR continuum frequency need to assume a dust temperature distribution in order to infer obscured SFR or dust mass. The left panel of Fig.\,\ref{fig2} shows the observed UV to IR (at $\lambda_0=160 \mu$m) flux ratio as a function of UV luminosity for UV or Ly$\alpha$ selected galaxies at $z\approx5-7$. The majority of measurements are upper limits and it is clear that dust continuum detections are only found in systems that are more UV-luminous. This likely points to a relatively monotonic relation between UV luminosity and (dust) mass. The most UV-luminous systems with the strongest constraints on UV-IR flux ratio are two luminous LAEs (CR7 and VR7). This indicates that those systems either have the least dust (at fixed UV luminosity), and/or the highest dust temperatures (at these frequencies, a higher dust temperature would decrease the UV-IR ratio for a fixed integrated IR luminosity). Such conditions are expected in LAEs, where high ISM ionisation states and little dust reddening are typical at $z\sim2$ (\cite[e.g. Trainor et al. 2016]{Trainor2016}).

Early [CII] observations of LAEs yielded contradictory results: studies found relatively high [CII] luminosities (\cite[Capak et al. 2015]{Capak2015}) or only strong upper limits on the [CII] luminosity (\cite[e.g. Ouchi et al. 2013]{Ouchi2013}), well below expectations from the local Universe. Other observations of UV-selected galaxies (but with known Ly$\alpha$ redshift) indicated mildly low [CII] luminosities (\cite[e.g. Pentericci et al. 2016]{Pentericci2016}). Weak [CII] emission in high-redshift galaxies severely challenged the promise of using ALMA as a `redshift-machine' and it was speculated that this was due to the strong Ly$\alpha$ emission (because of the pre-requisite of a known spectroscopic redshift) biasing follow-up observations towards galaxies with low metallicities and/or high ionisation states (both decreasing [CII] luminosity). 

More recently however, relatively strong [CII] emission has been detected in galaxies with strong Ly$\alpha$ emission (\cite[Matthee et al. 2017a; Carniani et al. 2018]{Matthee2017,Carniani2018}) and [CII] has also been detected in galaxies without strong Ly$\alpha$ emission (\cite[Smit et al. 2018]{Smit2018}). The right panel of Fig.\,\ref{fig2} shows that while there is a large range in [CII]-UV ratios in galaxies at $z\approx5-7$, relatively low [CII] luminosities are mostly found in faint galaxies (SFR$<25$ M$_{\odot}$ yr$^{-1}$), while [CII] is relatively strong in galaxies with higher SFRs. This indicates that either metallicity and/or ionisiation state are relatively strong functions of UV luminosity. As most LAEs are found in relatively faint galaxies, this explains why initial results on strong LAEs mostly found low [CII] luminosities, together with a better understanding of ALMA data reduction. Future deep observations still need to confirm these low [CII]/UV ratios.

\begin{figure}[h]
\begin{tabular}{cc}
 \includegraphics[height=5.85cm]{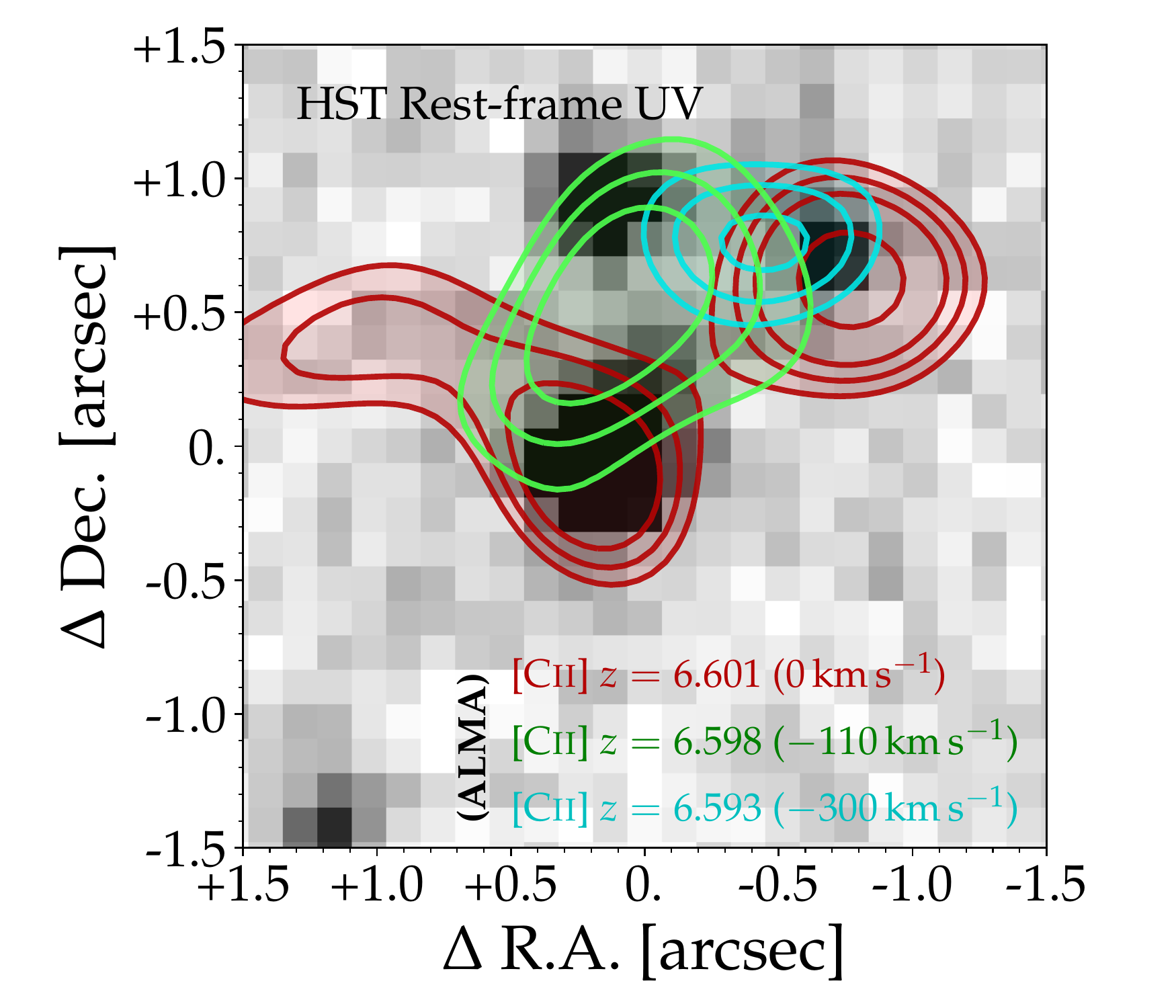} &
\hspace{-0.7cm} \includegraphics[height=5.85cm]{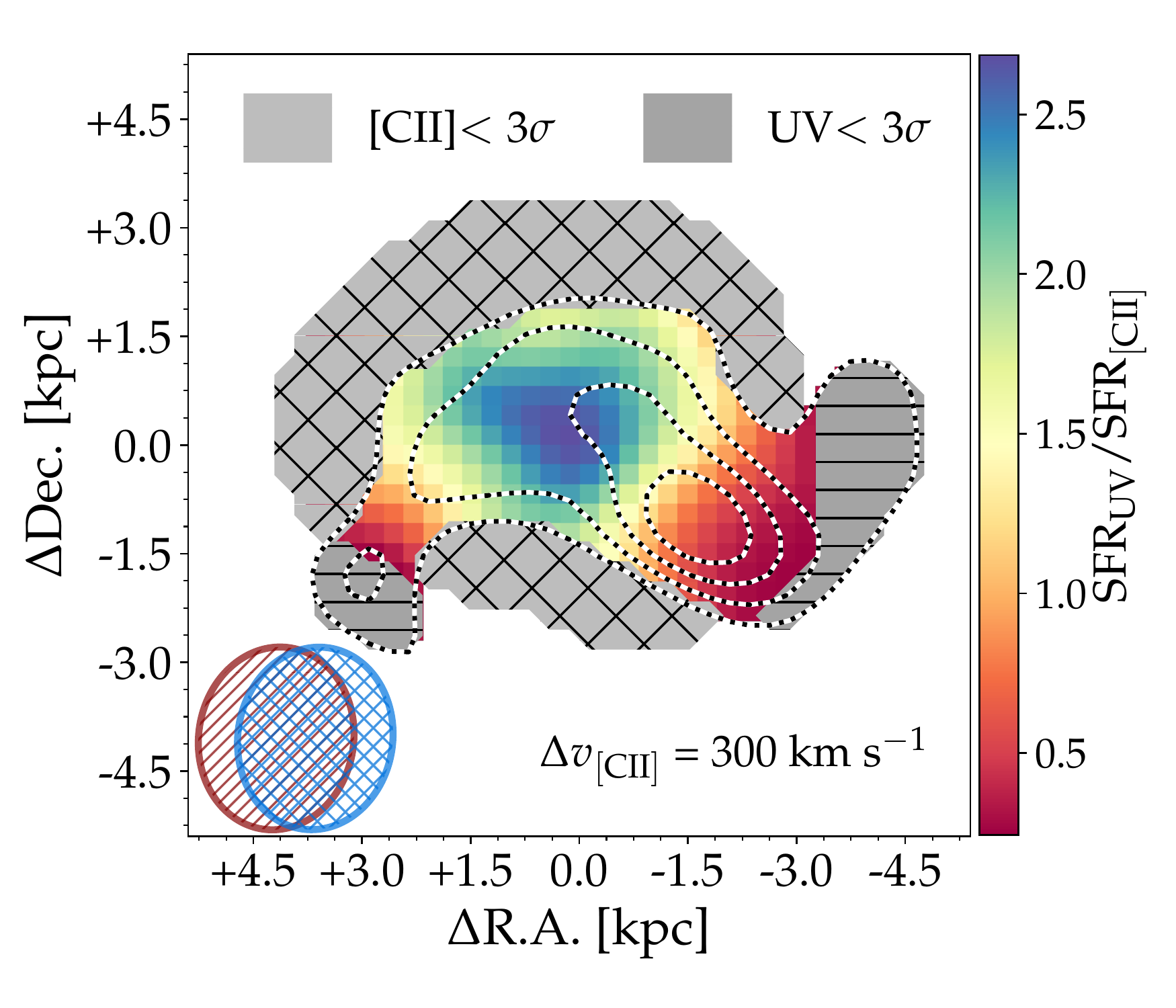}
 \end{tabular}
 \caption{Left: Resolved rest-frame UV and [CII] image of luminous LAE CR7 (\cite[Sobral et al. 2015, 2019]{Sobral2015,Sobral2019}) for which three UV and four [CII] components are identified. Right: Resolved UV to [CII] ratio of luminous LAE VR7 (Matthee et al. 2019). Variations in the UV-[CII] ratio of a factor 5 are observed on $\sim2$ kpc scales. }
   \label{fig4}
\end{figure}

Another important result is that [CII] emission in luminous LAEs typically consist of multiple components that are resolved in observations with $\sim2$kpc resolution (\cite[e.g. Matthee et al. 2017a; Carniani et al. 2018]{Matthee2017,Carniani2018}). Such components are seen with separations up to $\sim5$kpc and $\sim200$ km s$^{-1}$ and overlap (in most cases) with components seen in high-resolution rest-frame UV observations from {\it HST}, see Fig.\,\ref{fig4}. On small scales, [CII]-UV ratios can vary in the same galaxy, with differences up to factor 5 on $\sim2$kpc scales (Fig.\,\ref{fig4}; \cite[Matthee et al. 2019]{Matthee2019}). These observations indicate that luminous galaxies build-up through the relative complex assembly of different components.

\section{Conclusions \& Outlook}
We have highlighted recent results on deep, resolved spectroscopic observations of luminous LAEs at $z\sim6-7$. These reveal the first glimpse on the detailed properties of galaxies that reside in large ionised regions at the end of reionization. 
In the near future studies will likely increase the sample of observed galaxies and span the parameter space towards both fainter and brighter galaxies. Significant progress is anticipated when multiple emission lines in the same systems can be observed (for example with ALMA; e.g. \cite[Hashimoto et al. 2019]{Hashimoto2019}), but particularly with sensitive resolved infrared spectroscopy with {\it JWST}. In addition, the results on COLA1 show that deep, high resolution spectroscopy of the Ly$\alpha$ line is incredibly valuable, even in the era when spectroscopic confirmations can and (finally) will be provided using alternative emission lines. Ly$\alpha$ observations of high-redshift galaxies should therefore start to be seen even more as a physical tool in addition to be the means for spectroscopic redshift determination.


\end{document}